\documentclass{cmmse2011}

\usepackage{amssymb}
\usepackage{amssymb}
\usepackage{amsmath}
\usepackage{array} 
\usepackage{hyperref,bm,color}
\usepackage{graphicx}
\usepackage{multirow}
\usepackage{wrapfig}

\begin{document}




\markboth
  {An $O(N^3)$ approach for Hedin's $GW$ } 
  {Peter Koval, Dietrich Foerster, Daniel S\'anchez-Portal}


\title{An $O(N^3)$ implementation of Hedin's $GW$ approximation}



\author{Peter Koval}{koval.peter@gmail.com}{1}

\author{Dietrich Foerster}{d.foerster@cpmoh.u-bordeaux1.fr}{3}

\author{Daniel S\'anchez-Portal}{sqbsapod@sq.ehu.es}{1}


\affiliation{1}{Centro de F\'{\i}sica de Materiales CFM-MPC}
{Centro Mixto CSIC-UPV/EHU and DIPC, 
E-20018 San Sebasti\'an, Spain}

\affiliation{2}{CPMOH/LOMA}{University of Bordeaux, France}

\begin{abstract}
Organic electronics is a rapidly developing technology. 
Typically, the molecules involved in organic
electronics are made up of hundreds of atoms, 
prohibiting a theoretical description by wavefunction-based
ab-initio methods. Density-functional and Green's 
function type of methods scale less steeply with
the number of atoms. Therefore, they provide a suitable
framework for the theory of such large systems.

In this contribution, we describe an implementation, 
for molecules, of Hedin's $GW$ approximation.
The latter is the lowest order solution of a set of coupled 
integral equations for electronic
Green's and vertex functions that was found by Lars Hedin half 
a century ago.

Our implementation of Hedin's $GW$ approximation has two distinctive features:
i) it uses sets of localized functions to describe 
the spatial dependence of correlation functions, and 
ii) it uses spectral functions to treat their frequency dependence.
Using these features, we were able to achieve a 
favorable computational complexity of this approximation.
In our implementation, the number of operations grows 
as $N^3$ with the number of atoms $N$.

\end{abstract}

%
%
\section{Introduction}

\begin{wrapfigure}{c}{4.5cm}
\includegraphics[width=4.0cm,angle=0,clip]{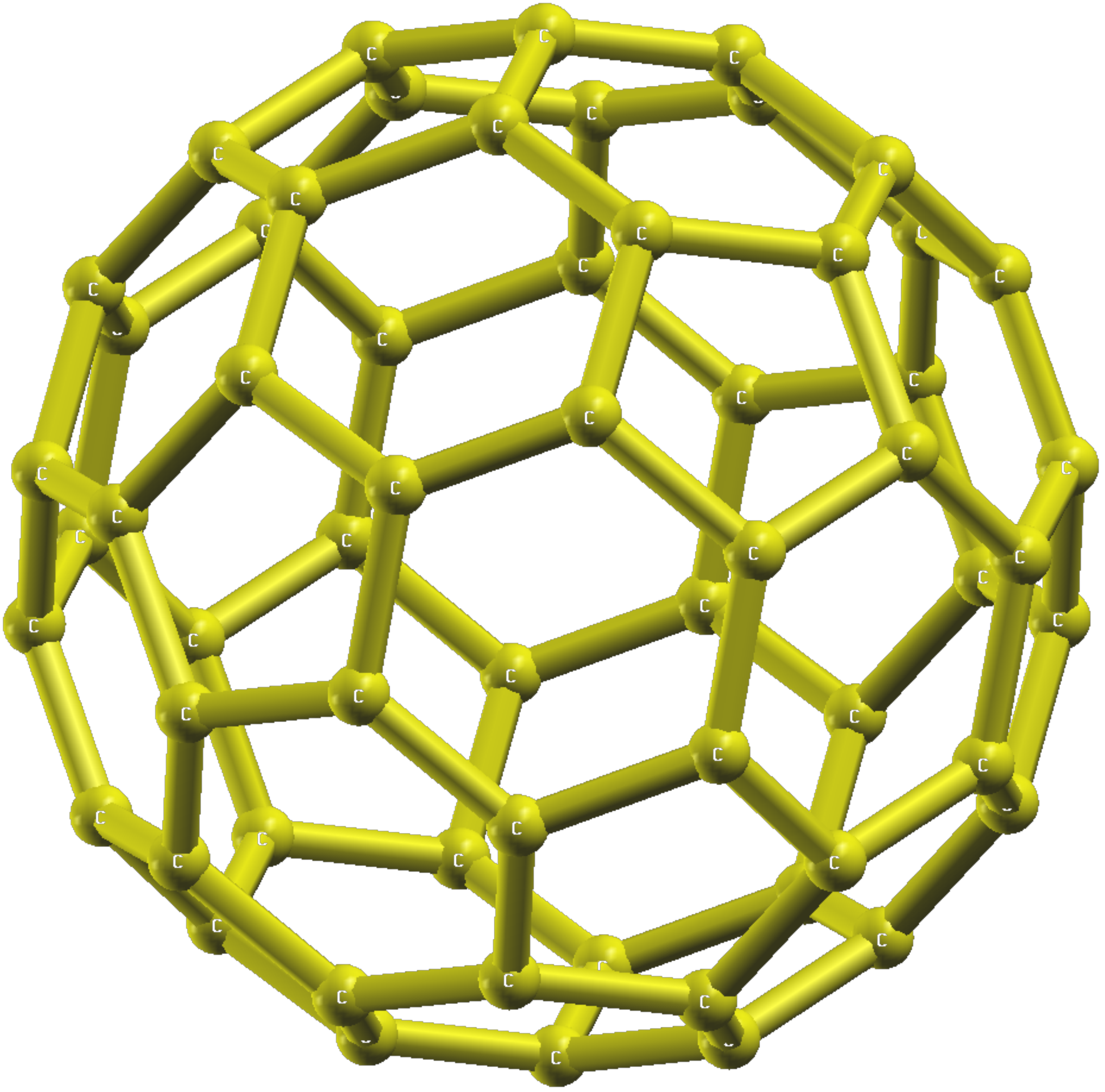}
\caption{Ball and stick model of fullerene C$_{60}$ 
produced with XCrysDen package \cite{xcrysden}.
\label{f:fullerene_geometry}}
\end{wrapfigure}

The promising field of organic electronics deals with large molecules 
of several tens or even hundreds of atoms~\cite{reviewOE}. 
For instance, fullerene C$_{60}$ is a frequently used subunit
in organic electronics and it alone consist of 60 atoms 
(see figure \ref{f:fullerene_geometry}).

Each individual molecule may be used in a device in many
different ways and there is an astronomically large number of different
promising molecules. 
As in many cases there is a limited knowledge of the relevant 
physical parameters, and it might be also interesting to explore
the potential of candidate molecules theoretically, before 
these molecules has been actually synthesised.

The geometry of large organic molecules can be reliably predicted by
density-functional theory (DFT)\cite{HK}. 
However, the properties of their excited
states such as the energy of the highest occupied (HOMO) and lowest
unoccupied molecular orbitals (LUMO), corresponding to adding and
subtracting one electron from the system respectively,
require a description of electronic correlations better than 
that provided by current functionals of DFT and its time-dependent
counterpart, TDDFT.

Such effects can be efficiently incorporated with the help of Hedin's method
that is based on Green's function. Hedin's $GW$ approximation for
one-electron Green's function is computationally cheaper than
wavefunction-based methods, although it remains computationally more
expensive than DFT and TDDFT within linear response.

The goal of our work is to develop a practical algorithm for Hedin's $GW$
approximation which is suitable for large organic molecules, allowing to
access the excited states of such molecules.

\section{ Theoretical framework for Hedin's $GW$ approximation}

\begin{wrapfigure}{c}{6.5cm}
\includegraphics[width=6.0cm,angle=0,clip]{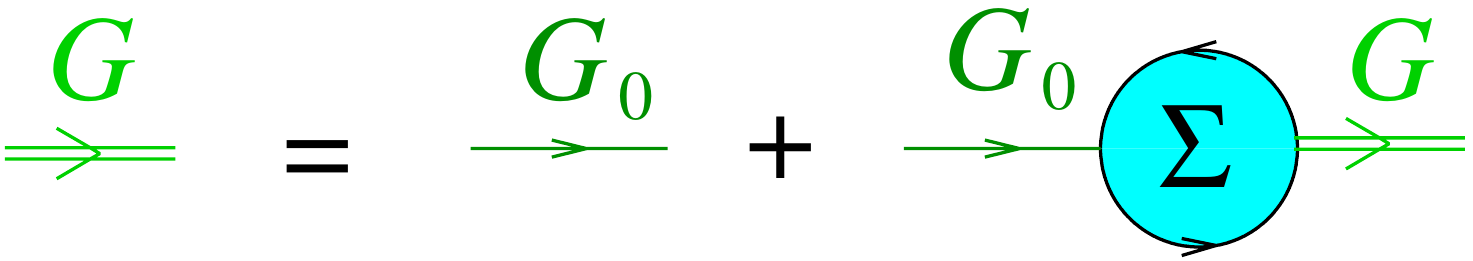}
\caption{Feynman diagram of Dyson equation (\ref{dyson_for_gf}).
\label{f:dyson_equation_diagram}}
\end{wrapfigure}Electronic Green's function (propagators) are useful
in condensed matter physics because many simple observables can be
computed in terms of them. At the same time, such Green's functions
remain simpler than many-body wavefunction.

Hedin's $GW$ is a useful approximation for the so-called
self-energy $\Sigma (\bm{r},\bm{r},\omega )$ that enters Dyson's equation
for an interacting electronic propagator $G(\bm{r},\bm{r}',\omega )$

\begin{equation}
G^{-1}(\bm{r}, \bm{r}',\omega) = G_0^{-1}(\bm{r}, \bm{r}',\omega)-\Sigma(\bm{r},\bm{r},\omega).
\label{dyson_for_gf}
\end{equation}%
Here, the inversions must be understood in operator sense
$\int G^{-1}(\bm{r}, \bm{r}'', \omega) G(\bm{r}'', \bm{r}', \omega)  dr'' = \delta(\bm{r}-\bm{r}')$
and  $G_{0}(\bm{r},\bm{r}^{\prime },\omega )$
stands for Green's function where electron-electron interactions
have been switched off.
It is obtained from an effective one-particle
Hamiltonian 
\begin{equation}
(\omega-H(\bm{r})) G_0(\bm{r},\bm{r}',\omega) = \delta(\bm{r}-\bm{r}').
\end{equation}

In this work we use a Kohn-Sham Hamiltonian~\cite{HK}, 
although Hartree-Fock
Hamiltonian also proves to be useful 
at this point~\cite{Blase}. 
Hedin's  $GW$
approximation for the self-energy $\Sigma(\bm{r},\bm{r},\omega)$ reads
\begin{equation}
\Sigma (\bm{r},\bm{r}',t) = \mathrm{i}G_0(\bm{r},\bm{r}',t)W_0(\bm{r},\bm{r}',t).
\label{self_energy}
\end{equation}
\begin{wrapfigure}[8]{c}{5.5cm}
\includegraphics[width=5.0cm,angle=0,clip]{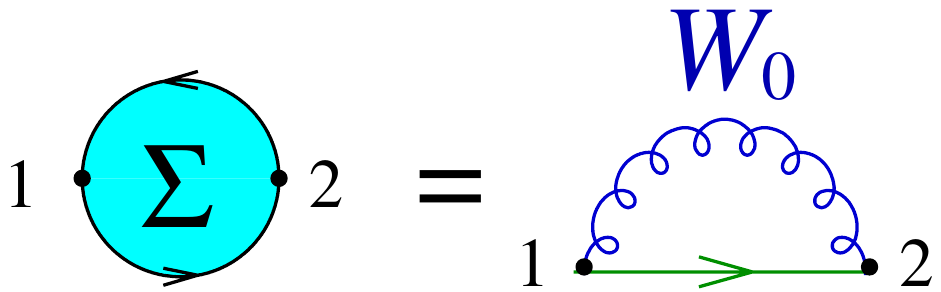}
\caption{Feynman diagram of self-energy (\ref{self_energy}).
\label{f:self_energy}}
\end{wrapfigure}
It involves the non interacting electronic Green's function %
$G_{0}(\bm{r},\bm{r}',t)$ and a screened Coulomb interaction $W_0(%
\bm{r},\bm{r}',t)$. This approximation is a solution of a truncated
version Hedin's equations \cite{Hedin,friedrich}. The name of this
approximation is taken from the simple form of the electronic self-energy
$\Sigma =\mathrm{i}GW$.

The screened Coulomb interaction $W_0$ can be easily calculated in frequency
domain using the so-called RPA approximation~\cite{RPA} 
\begin{equation}
W_0(\bm{r},\bm{r}',\omega ) =\left[\delta(\bm{r}-\bm{r}''')-v(\bm{r},\bm{r}'')
\chi_0(\bm{r}'',\bm{r}''',\omega )\right]^{-1}
v(\bm{r}''',\bm{r}'),
\label{scr_inter}
\end{equation}
\begin{wrapfigure}[7]{c}{9.0cm}
\includegraphics[width=8.5cm,angle=0,clip]{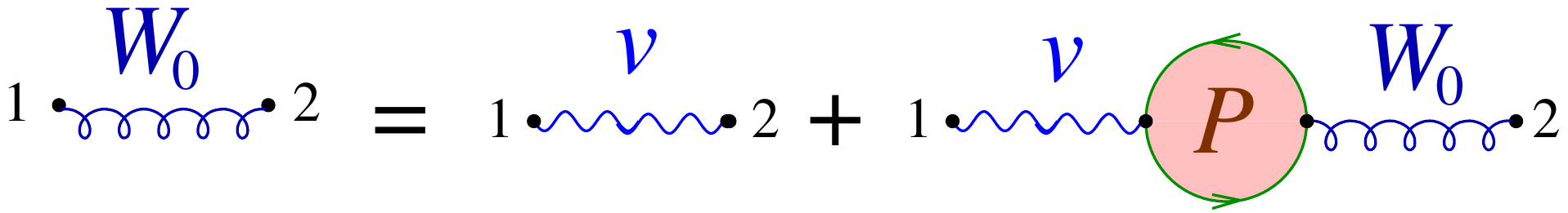}
\caption{Feynman diagram of screened Coulomb interaction (\ref{scr_inter}).
\label{f:screened_interaction}}
\end{wrapfigure}where $v(\bm{r},\bm{r}')\equiv |\bm{r}-\bm{r}'|^{-1}$ 
is the bare Coulomb interaction.
Here and in the following we 
assume integration over repeated spatial coordinates ($\bm{r}''
$ and $\bm{r}'''$ in equation (\ref{scr_inter})) on 
the right hand side of an equation if they do not appear on its left hand
side. The screened interaction (\ref{scr_inter}) is the sum of the 
bare Coulomb interaction created by a point charge at $\bm{r}'$, plus a 
correction due to the redistribution of charge induced in 
response to the total field \cite{RPA,friedrich}. 
The non-interacting response function $\chi_0(\bm{r},\bm{r}',t)$ is related to the
non-interacting Green's function
\begin{equation}
\mathrm{i}\chi_0(\bm{r},\bm{r}',t) =2 G_0(\bm{r},\bm{r}',t)G_0(\bm{r}',\bm{r},-t),
\label{noninter_response}
\end{equation}%
where a factor $2$ arises because of the summation over spin variable.

As we mentioned already, we construct the non-interacting Green's function
using an effective Kohn-Sham Hamiltonian~\cite{HK}
\begin{align}
H_{\mathrm{\mathrm{KS}}} &=-\frac{1}{2}\nabla^2+V_{\mathrm{KS}}, \\
V_{\mathrm{KS}} &=V_{\mathrm{ext}}+V_{\mathrm{Hartree}}+V_{\mathrm{xc}}
\text{, where }
V_{\mathrm{xc}}(\bm{r})=
\frac{\delta E_{\mathrm{xc}}}{\delta n(\bm{r})}.  \notag
\end{align}%
$E_{\mathrm{xc}}$ is a functional of the electronic density 
that includes the effects of exchange and correlation in an effective
way. Its functional derivative $V_{\mathrm{xc}}(\bm{r})$ is the
so-called exchange-correlation potential and it must be subtracted
from $\Sigma(\bm{r},\bm{r}',t)$ to 
avoid including the exchange-correlation interaction twice
in equation (\ref{self_energy}). This is accomplished with the substitution
\begin{equation*}
\Sigma(\bm{r},\bm{r}',t) \rightarrow \Sigma(\bm{r},\bm{r}',t) -
\delta(\bm{r}-\bm{r}')\delta(t) V_{\mathrm{xc}}(\bm{r})
\end{equation*}%
in Dyson's equation (\ref{dyson_for_gf}).

\section{A basis set of localized functions}

Having the equations (\ref{dyson_for_gf},\ref{self_energy},\ref{scr_inter},%
\ref{noninter_response}) at hand we introduce a basis set 
of localized functions
and rewrite the system of equations in the basis. We start with linear
combinations of atom orbitals (LCAO) to represent the 
non-interacting Green's function $G_0(\bm{r},\bm{r}',t)$
\begin{equation}
G_0(\bm{r},\bm{r}',t) = \sum_{ab} G^0_{ab}(t) f^{a}(\bm{r})f^{b}(\bm{r}'),
\label{g0_basis}
\end{equation}
where $f^{a}(\bm{r})$ are atom centered orbitals. The frequency (and time)
dependence has been factorized in the last equation. 
The treatment of the frequency (and time) 
dependence by spectral functions will be explained in 
section \ref{s:frequency-dependence}.
Inserting equation (\ref{g0_basis}) into the equation (\ref%
{noninter_response}), we obtain
\begin{equation}
\mathrm{i}\chi_0(\bm{r},\bm{r}',t) =2 
\sum_{abcd} G^0_{ab}(t) G^0_{cd}(-t)\, f^{a}(\bm{r})f^{d}(\bm{r})\, f^{b}(\bm{r}') f^{c}(\bm{r}').
\label{response_tensor}
\end{equation}
Products of localized orbitals such as 
$f^{a}(\bm{r})f^{d}(\bm{r})$ appear in the last equation. 
Although a product of localized orbitals is
also a localized function, such products do not form a suitable 
basis because
they contain many collinear functions. Several methods have been
proposed to 
construct more efficient basis to span 
the products of localized orbitals \cite{Blase,Casida,BaroniGW}. 
Here we use a basis of dominant products \cite{DF}
that is constructed individually for each atom pair. The dominant products
are identified as certain linear combinations of the original orbital products
and they are free of any collinearity within a given atom pair (with respect
to a given metric, here we have used the
Coulomb metric). Moreover, the original orbital products can be expressed
as linear combinations of dominant products 
\begin{equation}
f^{a}(\bm{r})f^{b}(\bm{r}) = V^{ab}_{\mu} F^{\mu}(\bm{r}).
\label{vertex_definition}
\end{equation}%
The three-index coefficient $V^{ab}_{\mu}$ will be referred to as the
\textit{product
vertex}. The product vertex is local or sparse by construction and indeed
the \textit{locality} of our construction is its main characteristic.

Considering Dyson's equation (\ref{dyson_for_gf}), we arrive at its tensor
counterpart
\begin{equation}
G_{ab}(\omega) = 
G^{0}_{ab}(\omega) + G_{aa'}(\omega)\Sigma^{a'b'}(\omega) G^{0}_{b'b}(\omega),
\label{dyson_tensor}
\end{equation}%
where matrix elements of the self-energy $\Sigma^{ab}(\omega)$ must be 
used
\begin{equation}
\Sigma^{ab}(\omega) = \iint f^{a}(\bm{r})\Sigma (\bm{r},\bm{r}',\omega) f^{b}(\bm{r}') \, d^3r d^3r'.
\end{equation}

Calculating the matrix elements of the self-energy by equation (\ref%
{self_energy}) and using (\ref{g0_basis}) for the non interacting Green's
function, we arrive at
\begin{equation}
\Sigma^{ab}(\omega)=\mathrm{i} \sum_{a'b'} G^0_{a'b'}(t) 
\int f^{a}(\bm{r})f^{a'}(\bm{r}) W_0(\bm{r},\bm{r}',t) f^{b'}(\bm{r}')f^{b}(\bm{r}')\, d^3r d^3r'.
\end{equation}
Using the identity (\ref{vertex_definition}), we rewrite the latter equation
as 
\begin{equation}
\Sigma^{ab}(\omega)=\mathrm{i} G^0_{a'b'}(t) V^{aa'}_{\mu} W_0^{\mu\nu}(t) V^{b'b}_{\nu},
\label{self_energy_tensor}
\end{equation}%
where the matrix elements of the screened Coulomb interaction appear
\begin{equation}
W_0^{\mu\nu}(t) = \iint  F^{\mu}(\bm{r}) W_0(\bm{r},\bm{r}',t) F^{\nu}(\bm{r}') \, d^3r d^3r'.
\end{equation}%

Finally, the equation (\ref{scr_inter}) gives rise to the corresponding
tensor expression
\begin{equation}
W_0^{\mu\nu}(\omega) = (\delta^{\mu}_{\nu'}-v^{\mu\mu'} \chi^0_{\mu'\nu'}(\omega))^{-1} v^{\nu'\nu}.
\label{scr_inter_tensor}
\end{equation}

The last expression can be elucidated by developing the operator $\left[
1-v\chi _{0}\right] ^{-1}$ in a geometric series
$\left[1-v\chi_0\right]^{-1} = 1+v\chi_0+v\chi_0v\chi_0+v\chi_0v\chi_0v\chi_0 \ldots$
The expressions (\ref{response_tensor}), (\ref{dyson_tensor}), (\ref%
{self_energy_tensor}) and (\ref{scr_inter_tensor}) are tensor counterparts
of Hedin's equations in coordinate space (\ref{noninter_response}), (\ref%
{dyson_for_gf}), (\ref{self_energy}) and (\ref{scr_inter}), correspondingly.
In the next section, we will present our method for treating the frequency
(and time) dependence of these tensor equations.

\section{Spectral function technique}

\label{s:frequency-dependence} Because of the discontinuities of 
the electronic
Green's functions, a direct, straightforward and accurate
computation of the response
function (\ref{response_tensor}) is practically impossible both in the time
domain and in the frequency domain. However, one can use an imaginary time
technique \cite{Godby} or spectral function representations to recover a
computationally feasible approach. In this work, we use spectral function
techniques and rewrite the time ordered operators as follows 
\begin{equation}
\begin{aligned}G^0_{ab}(t) &=
-\mathrm{i}\theta(t)\int_{0}^{\infty}ds\,\rho_{ab}^{+}(s)e^{-\mathrm{i}st}
+\mathrm{i}\theta(-t)\int_{-\infty}^{0}ds\,\rho_{ab}^{-}(s)e^{-\mathrm{i}st}; \\
\chi^0_{\mu\nu}(t) &=
-\mathrm{i}\theta(t)\int_{0}^{\infty}ds\,a_{\mu\nu}^{+}(s)e^{-\mathrm{i}st}
+\mathrm{i}\theta(-t)\int_{-\infty}^{0}ds\,a_{\mu\nu}^{-}(s)e^{-\mathrm{i}st}; \\
W_0^{\mu \nu }(t) &=
-\mathrm{i}\theta(t)\int_{0}^{\infty}ds\,\gamma_{+}^{\mu \nu}(s)e^{-\mathrm{i}st}
+\mathrm{i}\theta(-t)\int_{-\infty}^{0}ds\,\gamma_{-}^{\mu \nu}(s)e^{-\mathrm{i}st}; \\
\Sigma ^{ab}(t) &=
-\mathrm{i}\theta(t)\int_{0}^{\infty}ds\,\sigma_{+}^{ab}(s)e^{-\mathrm{i}st}
+\mathrm{i}\theta(-t)\int_{-\infty}^{0}ds\,\sigma_{-}^{ab}(s)e^{-\mathrm{i}st}, \\
\end{aligned}\label{spectral_1}
\end{equation}
where ``positive'' and ``negative''
spectral functions define the whole spectral 
function by means of Heaviside functions $\theta(t)$. For instance, the spectral
function of the electronic Green's function reads
$$
\rho_{ab}(s)=\theta(s)\rho^{+}_{ab}(s)+\theta(-s)\rho^{-}_{ab}(s).
$$
Transforming the first of equations (\ref{spectral_1}) to the frequency
domain, we obtain the familiar expression for the spectral representation of
a Green's function 
\begin{equation}
G^0_{ab}(\omega) = \int_{-\infty}^{\infty} \frac{\rho_{ab}(s) \, ds }{
\omega-s+\mathrm{i}\, \mathrm{sgn}(s) \varepsilon}.
\end{equation}%
Here $\varepsilon $ is a small line-broadening constant. 
In practice, the choice of $\varepsilon $ is related 
to the spectral resolution $\Delta \omega $ of the
numerical treatment.

As first application of representations (\ref{spectral_1}), we derive the
spectral function of the non interacting response $a_{\mu \nu }(s)$ using
equation (\ref{noninter_response}) as a starting point 
\begin{equation}
a_{\mu \nu }^{+}(s)=\iint V_{\mu }^{ab}\rho _{bc}^{+}(s_{1})V_{\nu
}^{cd}\rho _{da}^{-}(-s_{2})\delta (s_{1}+s_{2}-s)ds_{1}ds_{2}.
\label{sf_response_tensor}
\end{equation}%
Here, the convolution can be computed with fast Fourier methods and the
(time-ordered) response function $\chi _{\mu \nu }^{0}(\omega )$ can be
obtained with a Cauchy transformation 
\begin{equation}
\chi_{\mu \nu }^{0}(\omega )=\chi _{\mu \nu }^{+}(-\omega )+\chi _{\mu \nu
}^{+}(\omega ),\text{ where }\chi _{\mu \nu }^{+}(\omega )=\int_{0}^{\infty
}ds\,\frac{a_{\mu \nu }^{+}(s)}{\omega +\mathrm{i}\varepsilon -s}.
\label{sf2response}
\end{equation}

The calculation of the screened interaction $W_0^{\mu \nu }(\omega )$ must be
done with functions, rather than with spectral functions, because of the
inversion in equation (\ref{scr_inter_tensor}). The spectral function of the
screened interaction $\displaystyle\gamma ^{\mu \nu }(\omega )$ 
can be easily recovered from the screened interaction
itself \cite{friedrich}. Since 
$\mathrm{Im}\frac{1}{\omega +\mathrm{i}\varepsilon -s}$
is a representation of Dirac $\delta$-function when $\varepsilon$ goes
to zero, then
$\displaystyle\gamma ^{\mu \nu }(\omega )=-\frac{1}{\pi }\mathrm{Im}\,W_0^{\mu\nu}(\omega)$.

Deriving the spectral function $\sigma (\omega )$ of the self-energy, we
arrive at 
\begin{align}
\sigma _{+}^{ab}(s)& =\int_{0}^{\infty }\,\int_{0}^{\infty }\delta
(s_{1}+s_{2}-s)\,V_{\mu }^{aa^{\prime }}\rho _{a^{\prime }b^{\prime
}}^{+}(s_{1})V_{\nu }^{b^{\prime }b}\gamma _{+}^{\mu \nu
}(s_{2})ds_{1}ds_{2},  \label{spectral_3} \\
\sigma _{-}^{ab}(s)& =-\int_{-\infty }^{0}\,\int_{-\infty }^{0}\delta
(s_{1}+s_{2}-s)V_{\mu }^{aa^{\prime }}\rho _{a^{\prime }b^{\prime
}}^{-}(s_{1})V_{\nu }^{b^{\prime }b}\gamma _{-}^{\mu \nu
}(s_{2})ds_{1}ds_{2}.  \notag
\end{align}%
These expressions show that the spectral function of a convolution is given
by a convolution of the corresponding spectral functions.

\subsection{Discretization of frequency-dependent quantities}

The spectral functions in equation (\ref{sf_response_tensor}) are merely a
set of poles at (eigen)energies $E$ 
\begin{equation}
\rho _{ab}^{+}(\omega )=\sum_{E>0}\delta (\omega -E)X_{a}^{E}X_{b}^{E},\
\rho _{ab}^{-}(\omega )=\sum_{E<0}\delta (\omega -E)X_{a}^{E}X_{b}^{E}.
\label{sf_0}
\end{equation}%
Here the eigenvectors $X_{a}^{E}$ diagonalize the corresponding Kohn-Sham
Hamiltonian $$H^{ab}X_{b}^{E}=ES^{ab}X_{b}^{E},$$
where the Hamiltonian and the overlap matrices of 
atomic orbitals $f^{a}(\bm{r})$ enter 
\begin{equation}
H^{ab}=\int f^{a}(\bm{r})H(\bm{r})f^{b}(\bm{r})d^{3}r,\text{ and }%
S^{ab}=\int f^{a}(\bm{r})f^{b}(\bm{r})d^{3}r.
\end{equation}%
In practice, we use the SIESTA package \cite{siesta} that gives the orbitals 
$f^{a}(\bm{r})$, eigenvectors $X_{a}^{E}$ and eigenvalues $E$ 
for a given molecule as the
output of a DFT calculation.

The use of fast Fourier techniques for convolution, 
for instance in equation (\ref{sf_response_tensor}),
requires that the spectral functions $\rho _{bc}^{+}(\omega )$, $\rho%
_{da}^{-}(\omega )$ be known at equidistant grid points $\omega _{j}=j\Delta
\omega ,j=-N_{\omega }\ldots N_{\omega }$, rather than at a set of energies
resulting from a diagonalization procedure. The solution for this problem
(discretization of spike-like functions) is known and well tested \cite%
{Shishkin-Kresse:2006}. We define a grid of points that covers the
whole range of eigen energies $E$. Going through the poles $E$, we assign
their spectral weight $X_{a}^{E}X_{b}^{E}$ to the neighboring grid points $n$
and $n+1$ such that $\omega _{n}\leq E<\omega _{n+1}$ according to the
distance between the pole and the grid points 
$\displaystyle p_{n,\,ab}=\frac{\omega _{n+1}-E}{\Delta \omega }X_{a}^{E}X_{b}^{E},\
p_{n+1,\,ab}=1-p_{n,\,ab}.$
Such a discretization keeps both the spectral weight and the center of mass
of a pole. It also reduces the number of operations that are needed to
calculate the non interacting response function $\chi _{\mu \nu }^{0}(\omega%
)$. This is so because the number of frequencies $N_{\omega }$ can be kept
small (typically a few hundred points) even for large molecules while the
number of states $N_{\mathrm{orb}}$ grows linearly with the size of the
system.

\subsection{The second window technique}

The discretization of spectral weight helps to control the computational
complexity for large molecules. However, we are actually interested in the
properties of low lying levels (HOMO and LUMO and several levels below and
above). At first sight one might think that one could neglect the
contributions of high energy spectral weights in the Cauchy transformation.
However, neglecting the high energy spectral weight actually results in a
wrong real part of the functions. Fortunately, the high energy spectral
weight tolerates a coarser grid \cite{df-pk-dsp:2011}. Therefore, we
calculate each spectral function twice: once with a higher resolution in a
low frequency range, and a second time with a lower resolution but in the
whole range. The Cauchy transformation for such a two-window representation
must be modified as follows

\begin{multline}
\chi _{\mu \nu }^{0}(\omega +\mathrm{i}\varepsilon _{\text{small}%
})=\int_{-\lambda }^{\lambda }ds\frac{a_{\mu \nu }(s)}{\omega +\mathrm{i}%
\varepsilon _{\text{small}}-s}\ +\left( \int_{-\Lambda }^{-\lambda
}+\int_{\lambda }^{\Lambda }\right) ds\frac{a_{\mu \nu }(s)}{\omega +\mathrm{%
i}\varepsilon _{\text{large}}-s} \\
=\chi _{\mu \nu }^{\text{small window}}(\omega +\mathrm{i}\varepsilon _{%
\text{small}})+\left[ \chi _{\mu \nu }^{\text{large window}}(\omega +\mathrm{%
i}\varepsilon _{\text{large}})\right] _{\text{truncated spectral function}}.
\label{win2-for-response}
\end{multline}%
After the calculation of spectral functions in both windows, we truncate the
spectral function in the second window in the range $0\ldots \lambda $, do
Cauchy transform of both spectral functions and update (by a linear
interpolating procedure) the function in the first window with the truncated
function from the second window.

We use the second window technique both for the non interacting response
function $\chi _{\mu \nu }^{0}(\omega )$ and for the self-energy $%
\Sigma^{ab}(\omega)$.

\section{Non-local compression of the product basis}

\label{s:non_local}

The basis of dominant products is optimal within a given atom pair, but
unfortunately, there is still a lot of collinearity between dominant products
belonging to different pairs. This collinearity is an indication 
that the size of the product basis can be strongly reduced.
Even for the molecules of modest size considered in 
Section~\ref{ss:hydrocarbons} 
the basis set of dominant product becomes so large that
hampers the storage of the (non-interacting) response function (\ref%
{sf2response}) and slows down the inversion in the calculation of the
screened interaction (\ref{scr_inter_tensor}). In order to improve the
situation we perform a non-local (global) contraction of the basis 
of dominant product.  We start by considering a sum-over-states 
expression for the non-interacting
response function in the basis of dominant products

\begin{equation}
\chi _{\mu \nu }^{0}(\omega )=2\sum_{E,F}V_{\mu }^{EF}\frac{n_{F}-n_{E}}{%
\omega +i\varepsilon -(E-F)}V_{\nu }^{EF}\text{, where }
V_{\mu}^{EF}=X_{a}^{E}V_{\mu }^{ab}X_{b}^{F}.
\end{equation}%
The response $\chi _{\mu \nu }^{0}(\omega )$ is built up from vectors $%
V_{\mu }^{EF}$ that represent electron-hole pair excitations. 
One can use these vectors to identify important directions in
the space of dominant products. The number of electron--hole pairs $EF$
grows as $N^{2}$ with the molecular size while the dimension of dominant
product basis is $O(N)$ by construction (due to the localization of the basis
orbitals). Therefore, one has to limit the set
of electron-hole pairs $EF$ from the beginning to keep the efficiency 
of the algorithm, particularly if one uses a
diagonalization-based procedure for generating the (globally) optimal basis.
Because of the inherent limitations of LCAO to represent high energy features, 
and the fact that we are mainly interested in the lowest lying excitations, 
we choose $O(N)$ low-energy electron-hole pairs 
\begin{equation}
\{X_{\mu }^{n}\}\equiv \text{subset of }\{V_{\mu }^{EF}\}\text{ limited by }%
|E-F|<E_{\mathrm{threshold}},n=1\ldots N_{\mathrm{rank}}.
\label{subsetof_vef}
\end{equation}%
After the initial selection according to the energy criterion $|E-F|<E_{%
\mathrm{threshold}}$, we define a metric $g^{mn}$ 
\begin{equation}
g^{mn}=X_{\mu }^{m}v^{\mu \nu }X_{\nu }^{n},\text{ where }v^{\mu \nu }=\iint
F^{\mu }(\bm{r})|\bm{r}-\bm{r}^{\prime }|^{-1}F^{\mu }(\bm{r}^{\prime
})d^{3}rd^{3}r^{\prime }.
\label{projector0}
\end{equation}%
After diagonalizing the metric $g^{mn}\xi _{n}^{\lambda }=\lambda \xi
_{m}^{\lambda }$, we can identify important directions (like in the
construction of the basis of dominant products \cite{DF,Aryasetiawan}) by
building linear combinations of the original vectors $X_{\mu }^{m}$ and by
choosing only eigenvectors with eigenvalues above a suitable threshold value 
\begin{equation}
Z_{\mu }^{\lambda }\equiv X_{\mu }^{m}\xi _{m}^{\lambda }/\sqrt{\lambda }.
\label{orthogonal}
\end{equation}%
These linear combinations can be used to expand the original response
function $\chi _{\mu \nu }^{0}(\omega )$ in terms of fewer functions 
\begin{equation}
\chi _{\mu \nu }^{0}(\omega )=Z_{\mu }^{m}\chi _{mn}^{0}(\omega )Z_{\nu
}^{n}.  \label{expand_in_the_basis}
\end{equation}%
In order to express $\chi _{mn}^{0}(\omega )$ in terms of $\chi _{\mu \nu
}^{0}(\omega )$ we multiply equation (\ref{expand_in_the_basis}) with $%
Z_{\mu }^{m}v^{\mu \nu }$ from both sides and notice that $Z_{\mu%
}^{m}v^{\mu \nu }Z_{\nu }^{n}\equiv Z_{\mu }^{m}Z_{n}^{\mu }=\delta _{n}^{m}$%
. Therefore, the response function can be ``%
compressed'' by using basis vectors $Z_{n}^{\nu }\equiv v^{\mu \nu }Z_{\nu }^{n}$ 
\begin{equation}
\chi _{mn}^{0}(\omega )=Z_{m}^{\mu }\chi _{\mu \nu }^{0}(\omega )Z_{n}^{\nu}.
\label{response_compressed}
\end{equation}

The particular choice of the Coulomb metric $v^{\mu \nu }$ in equation 
(\ref{projector0}) simplifies 
the computation of the Coulomb screened interaction (\ref{scr_inter_tensor}).
We can rewrite the equation (\ref{scr_inter_tensor}) in
terms of a Taylor series 
\begin{equation}
W_0^{\mu \nu }=v^{\mu \nu }+v^{\mu \mu ^{\prime }}\chi _{\mu ^{\prime }\nu
^{\prime }}^{0}v^{\nu ^{\prime }\nu }+v^{\mu \mu ^{\prime }}\chi _{\mu
^{\prime }\nu ^{\prime }}^{0}v^{\nu ^{\prime }\mu ^{\prime \prime }}\chi
_{\mu ^{\prime \prime }\nu ^{\prime \prime }}^{0}v^{\nu ^{\prime \prime }\nu
}+\cdots   \label{W-series}
\end{equation}%
Inserting here the response function $\chi _{\mu \nu }^{0}$ according to
equation (\ref{expand_in_the_basis}) and recalling the identity $Z_{\mu
}^{m}Z_{n}^{\mu }=\delta _{n}^{m}$, one arrives at 
\begin{align}
W_0^{\mu \nu }& =v^{\mu \nu }+Z_{m}^{\mu }\chi _{mr}^{0}\left[ \delta
_{rn}+\chi _{rn}^{0}+\chi _{rs}^{0}\chi _{sn}^{0}+\cdots \right] Z_{n}^{\nu
}=  \label{chi_RPA} \\
& =v^{\mu \nu }+Z_{m}^{\mu }\chi _{mn}^{\mathrm{RPA}}Z_{n}^{\nu },\text{
where }\chi _{mn}^{\mathrm{RPA}}\equiv \left( \delta _{mk}-\chi
_{mk}^{0}\right) ^{-1}\chi _{kn}^{0}.  \notag
\end{align}

At this point it should be also be noted that the self-energy $\Sigma
_{x}^{ab}(\omega )$ that corresponds to the instantaneous part of the
screened interaction $v^{\mu \nu }$ is computed separately \cite%
{friedrich,df-pk-dsp:2011} without any non local compression.

\section{Computational complexity of the algorithm}

The number of mathematical operations spent in different parts of the
approach presented above can be estimated if the dimensions of the
corresponding matrices are known. The numbers that determine the complexity
of the algorithm are the number of atomic orbitals $N_{\mathrm{orb}}$, the
number of dominant functions $N_{\mathrm{prod}}$ and the number of
frequencies $N_{\omega }$. The number of orbitals and the number of dominant
products are proportional to the number of atoms $N$ by construction. The
number of frequencies affects the run time linearly, but it is independent
of the number of atoms. The non-local basis of section \ref{s:non_local},
can be constructed in $O(N^{3})$ operations because $N_{\mathrm{rank}}$ in
equation (\ref{subsetof_vef}) can be kept proportional to number of orbitals.
In practical calculations we have found that converged results are achieved with
$N_{\mathrm{rank}}\sim 5 N_{\mathrm{orb}}$. For large molecules, the number
of important eigenvectors $N_{\mathrm{subrank}}$ after dropping small
eigenvalues $\lambda $ in equation (\ref{orthogonal}) is approximately $N_{%
\mathrm{orb}}$.
No part of the algorithm scales worse than $O(N^{3})$ \cite{df-pk-dsp:2011}.
There are several portions of the code where $O(N^{3})$ operations are
needed. However, only two of them have an appreciable impact on the 
run time: the computation of the response function and the computation
of the self-energy. 
Both of them scale as 
$O(N_{\mathrm{prod}}^{2}N_{\mathrm{subrank}}N_{\omega})$
and give rise to an overall $O(N^{3})$ scaling of the run time.

\section{Applications to organic molecules}

The methods described in the previous sections were carefully tested on
several molecules. In this paper, we present two examples: calculations of
HOMO and LUMO levels of three aromatic hydrocarbons (benzene, naphthalene
and anthracene) and a calculation of the HOMO and LUMO levels of fullerene C$%
_{60}$.

\subsection{Aromatic hydrocarbons}

\label{ss:hydrocarbons}

From the interacting Green's function $G_{ab}(\omega )$ we calculate the
density of states (DOS)
$\rho (\omega )=-S^{ab}\mathrm{Im}G_{ab}(\omega )/\pi$
and we then determine the positions of the HOMO and LUMO levels from the DOS.

\begin{wraptable}{c}{6.8cm}
\renewcommand{\tabcolsep}{0.1cm}
\begin{tabular}[t]{|p{3.2cm}|p{1.3cm}|p{1.4cm}|}
\hline
\hfil Picture \hfil  & IP, eV   & EA, eV  \\
\hline
{\hfil\includegraphics[width=1.5cm, angle=0,clip]{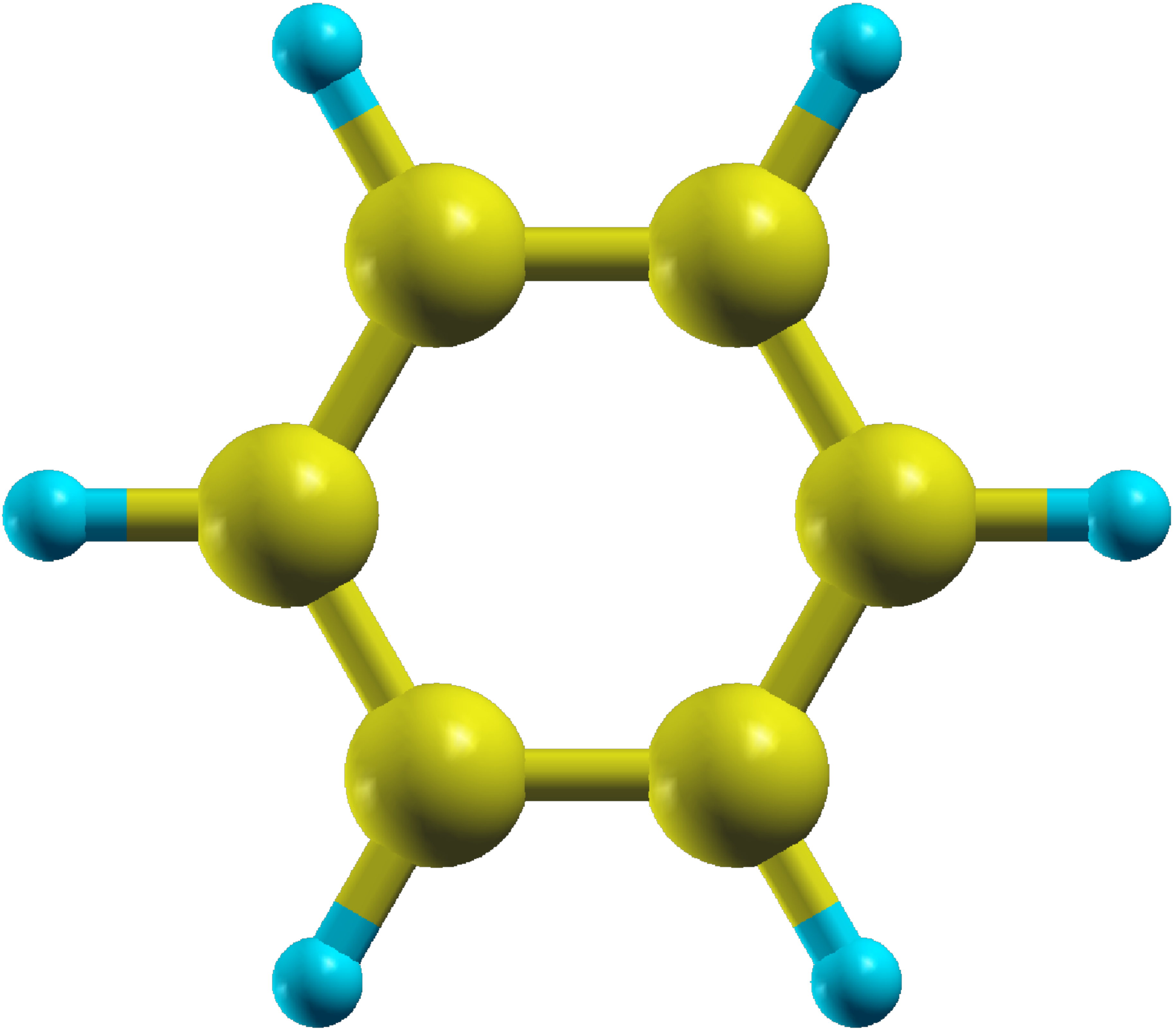}\hfil}   
                     & \vspace{-1.2cm} 8.82 \newline (9.25) 
                                & \vspace{-1.2cm}-1.43 \newline (-1.12)  \\
{\hfil\includegraphics[width=2cm, angle=0,clip]{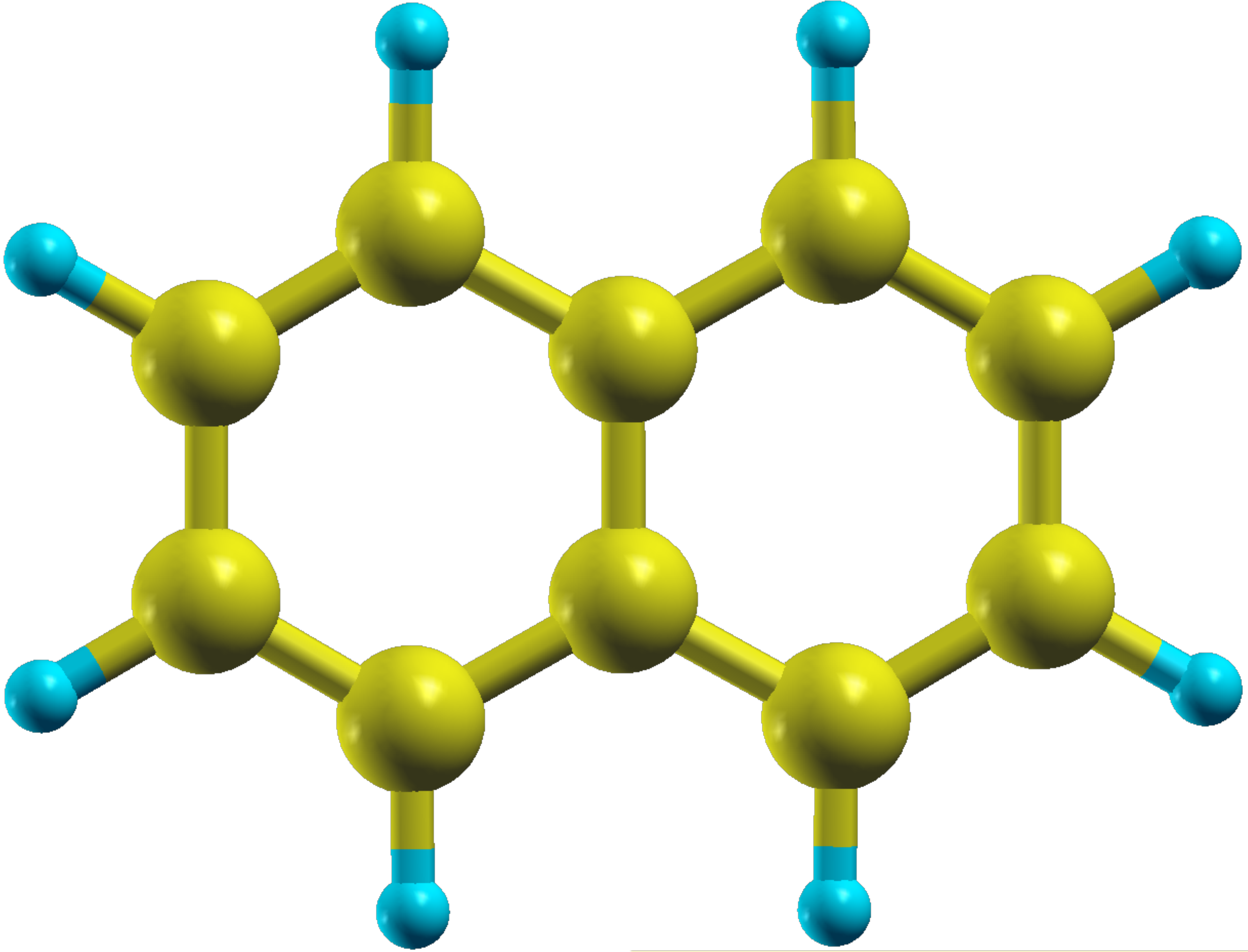}\hfil} 
                     & \vspace{-1.2cm} 7.58 \newline (8.14)  
                                & \vspace{-1.2cm} -0.15 \newline (-0.19) \\
{\hfil\includegraphics[width=3cm, angle=0,clip]{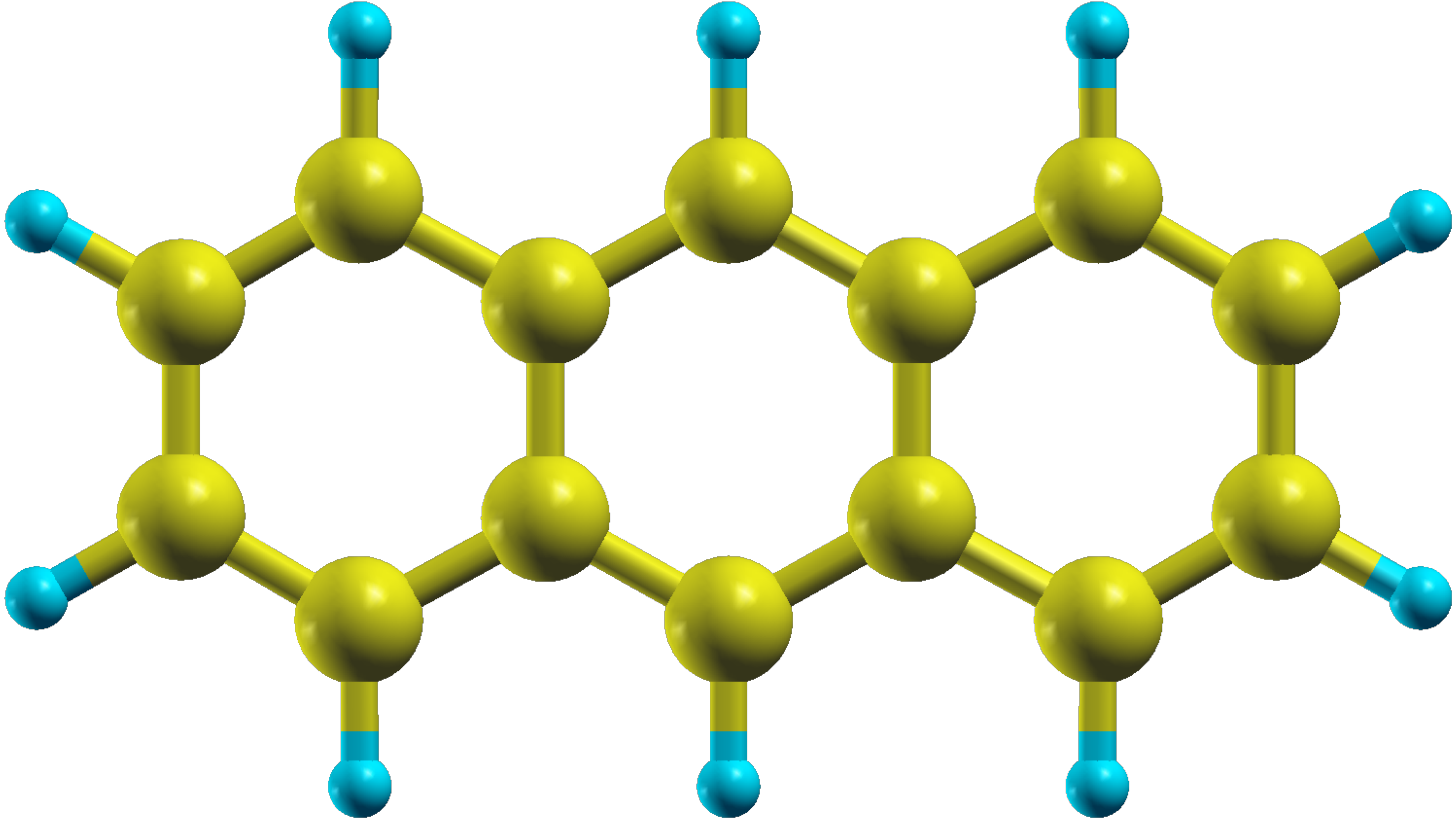}\hfil} 
                     &  \vspace{-1.2cm} 6.87 \newline (7.44)        
                                & \vspace{-1.2cm} 0.73 \newline (0.530)  \\
\hline
\end{tabular}
\caption{\label{t:homo_lumo_hydrocarbons} The ionization potentials (IP)
and electron affinities (EA) of benzene, naphthalene and anthracene.
Experimental data \cite{experimental_data} are given in brackets.}
\end{wraptable}

The results of this procedure for aromatic hydrocarbons
are collected in table \ref{t:homo_lumo_hydrocarbons}. One can see that our
(LDA+$G_0W_0$) approach delivers \textit{qualitatively correct predictions} for
the ionization potentials (IP) and electron affinities (EA) of benzene and
naphthalene (donors) and anthracene (acceptor). On the other hand, the LUMO
of the underlying DFT calculation is always below the vacuum level. The
calculations have been done on top of DFT-SIESTA calculations.
We used 
pseudo potentials of Troullier-Martins type \cite{Troullier-Martins} and
the Perdew-Zunger exchange-correlation functional \cite{Perdew-Zunger}. 
We found
that rather extended atomic orbitals must be used to achieve converged
results in our $GW$ approach. The energy shift parameter \cite%
{Artacho-about-Energy-Shift}, that controls the spatial extension of atomic
orbitals has been set to 3~meV for benzene, and to  20~meV for naphthalene
and anthracene. The spectral functions have been discretized in two energy
windows, with each window containing $N_{\omega }=64$ frequency points.

\subsection{Fullerene C$_{60}$}

\begin{wraptable}{c}{6cm}
\renewcommand{\tabcolsep}{0.1cm}
\vspace{-1.0cm}
\begin{tabular}{|l|c|c|}
\hline
 Source       & IP, eV   & EA, eV   \\
\hline
 Our LDA+$G_0W_0$ & 7.33     & 2.97    \\
 Experimental \cite{experimental_data} 
                  & 7.58     & 2.65   \\
\hline
\end{tabular} 
\caption{\label{t_c60}The IP and EA of fullerene C$_{60}$ calculated
with our method and corresponding experimental data.}
\end{wraptable}The fullerene C$_{60}$ and its derivatives are very popular
ingredients in organic semiconductors and extensive experimental data and
theoretical computations are available for the basic fullerene. We found
that our LDA+$G_0 W_0$ results are in very good agreement with
experimental data (see table \ref{t_c60}). The computational parameters of
this calculation are the same as in subsection \ref{ss:hydrocarbons}, while
the energy shift parameter is chosen to be 15~meV. The number of frequency
points was chosen rather large $N_{\omega }=128$ and the calculation has
been done with 8 cores of a Nehalem machine (Intel\textregistered E5520 
\@ 2.27GHz, Cache 8M/DDR3 RAM 24GB). The current version of the code consumed
26 hours of wall clock time. 

\begin{wrapfigure}[14]{c}{7cm}
\includegraphics[width=6.5cm,angle=0,viewport=50 50 400 300,clip]{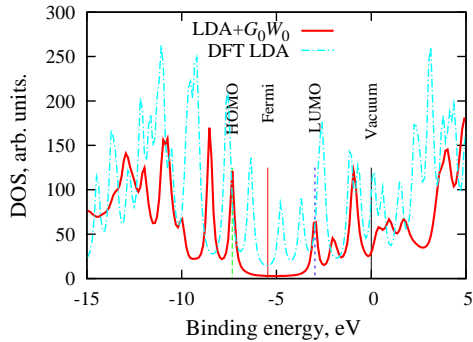}
\caption{DOS of fullerene C$_{60}$ computed with our LDA-$G_0 W_0$ approach.
\label{f:fullerene_dos}}
\end{wrapfigure}

A comparison of DOS calculated with DFT LDA Hamiltonian and with our LDA+$G_0 W_0$
approach is shown in figure \ref{f:fullerene_dos}. Such a result 
is a typical when Hedin's $GW$ approach is applied on top of a LDA calculation.
$GW$ HOMO has lower energy than DFT HOMO. Therefore, the change density $n(\bm{r})$
will be more localized in the $GW$ calculation. $GW$ LUMO has higher energy than DFT LUMO.
Therefore, the change density $n(\bm{r})$ will be more delocalized in the $GW$ calculation.

\section{Conclusions}

We have described our approach to Hedin's $GW$ approximation for finite
systems. This approach allows to compute the interacting Green's function on
a frequency grid. The density of states is our output and it provides HOMO
and LUMO levels in reasonable agreement with experiment. The complexity of
the approach scales with the third power of the number of atoms, while the
needed memory scales with the second power of the number of
atoms. These features make our
approach suitable for treating the large molecules that are used in organic
semiconductors. 

\section*{Acknowledgments}

We thank James Talman for inspiring discussions, correspondence and
essential algorithms and computer codes \cite{Talman}, that are used in
our implementation. We are indebted to the organizers of the ETSF2010
meeting at Berlin for feedback and perspective on the ideas of this paper.
Arno Schindlmayr, Xavier Blase and Michael Rohlfing helped with extensive
correspondence on various aspects of the $GW$ method. DSP and PK acknowledge
financial support from the Consejo Superior de Investigaciones Cient\'{\i}%
ficas (CSIC), the Basque Departamento de Educaci\'on, UPV/EHU (Grant No.
IT-366-07), the Spanish Ministerio de Ciencia e Innovaci\'on (Grants No.
FIS2010-19609-C02-02) and, the ETORTEK program
funded by the Basque Departamento de Industria and the Diputaci\'on Foral de
Guipuzcoa.

%
%

\end{document}